\documentclass[12pt]{iopart}
\usepackage{graphicx}
\expandafter\let\csname equation*\endcsname\relax
\expandafter\let\csname endequation*\endcsname\relax
\usepackage{amsmath}
\usepackage{iopams}
\usepackage{amssymb}
\usepackage{color}
\newcommand {\fabs}[1] {\left| #1 \right|}

\newcommand {\cU}{{\cal{U}}}

\newcommand{\ket}[1]{\ensuremath{|#1\rangle}}
\newcommand{\bra}[1]{\langle#1|}
\newcommand{\braket}[2]{\langle#1|#2\rangle}
\newcommand{\ketbra}[2]{|#1\rangle\langle#1|}

\usepackage{ulem}

\begin{document}
\title[Creating exotic angular momentum states by shaking an optical lattice]{Shaken not stirred:\\ Creating exotic angular momentum states by shaking an optical lattice}
\author{Anthony Kiely}
\ead{anthony.kiely@umail.ucc.ie}
\address{Department of Physics, University College Cork, Cork, Ireland}

\author{Albert Benseny}
\address{Quantum Systems Unit, Okinawa Institute of Science and Technology Graduate University, 904-0495 Okinawa, Japan}

\author{Thomas Busch}
\address{Quantum Systems Unit, Okinawa Institute of Science and Technology Graduate University, 904-0495 Okinawa, Japan}

\author{Andreas Ruschhaupt}
\address{Department of Physics, University College Cork, Cork, Ireland}
%
%
\begin{abstract}
We propose a method to create higher orbital states of ultracold atoms in the Mott regime of an optical lattice.
This is done by periodically modulating the position of the trap minima (known as shaking) and controlling the interference term of the lasers creating the lattice. These methods are combined with techniques of shortcuts to adiabaticity. As an example of this, we show specifically how to create an anti--ferromagnetic type ordering of angular momentum states of atoms. The specific pulse sequences are designed using Lewis--Riesenfeld invariants and a four--level model for each well.
The results are compared with numerical simulations of the full Schr\"{o}dinger equation.  
\end{abstract}
\maketitle


\section{Introduction}

Optical lattices have proven to be highly versatile systems for investigating quantum many body physics \cite{bloch_2008, lewenstein_2012}.
A notable example of this is the observation of the phase
transition between a superfluid and a Mott--insulator state
\cite{mott_fluid,mott_fluid_cs}.
These results are achieved with atoms trapped in the lowest band of the optical lattice.
%
However, in the solid state, the orbital degree of freedom also plays an important role in many of the complex phases.
For instance, many models in high temperature superconductivity involve higher orbital occupations \cite{Kamihara:06,Luke:98,Ohtomo:04}.
As a result, there has been a lot of interest recently in the physics of higher bands of optical lattices \cite{lewenstein_2011, higher_bands_review}.
The bosonic Hubbard model describing the lowest band has been extended to incorporate higher Bloch bands \cite{isacsson_2005}
and Bose-Einstein condensation with nonzero orbital momenta has been studied \cite{kuklov_2006, liu_2006}.
Many exotic phases have been predicted to occur due to the interplay of interactions and the higher bands \cite{hebert_2013}.

Recently, first experiments have been performed realising multiorbital systems with ultracold atoms \cite{browaeys_2005, mueller_2007} where the lifetimes of atoms in the excited state have been long enough to observe tunnelling dynamics. In particular, the formation of a superfluid in the higher bands has been experimentally achieved \cite{wirth_2011}.
The condensate formation in the higher bands has been used to investigate topologically induced avoided band crossing \cite{oelschlaeger_2012}.

Engineering quantum states in higher bands is therefore of large interest and several techniques have been developed to manipulate the state of atoms in an optical
lattice \cite{higher_bands_review}.
%
One example of this is periodic modulation of the lattice amplitudes in order to
induce controlled transitions to higher orbital states \cite{sowinski_2012} or transitions to motional eigenstates \cite{holder_2007}.
Higher orbitals have also be generated by stimulated Raman transitions \cite{mueller_2007}.

Another possibility is to shake the lattice in one direction, i.e., a periodic modulation of the position of the trap minima.
The idea of shaking a single trap has been previously used for a variety of other tasks such as vibrational state inversion of a condensate in a trap \cite{buecker_2013} and Ramsey interferometry using the motional states of the condensate \cite{frank_2014}.
Shaking of an optical lattice in one direction has been examined theoretically in connection with quantum
computation \cite{schneider_2012} and especially to create higher orbital states in the lattice
\cite{zhang_2014,  strater_2015, zhang_2015}. 
The latter has also been realised experimentally \cite{parker_2013, khamehchi_2016}.
Recently there has
been work which combines both amplitude and position modulation of the lattice potential
using optimal control in order to transfer atoms between different vibrational states~\cite{hallaji_a2015}. 

\begin{figure}[t]
\begin{center}
\includegraphics[angle=0,width=0.45\linewidth]{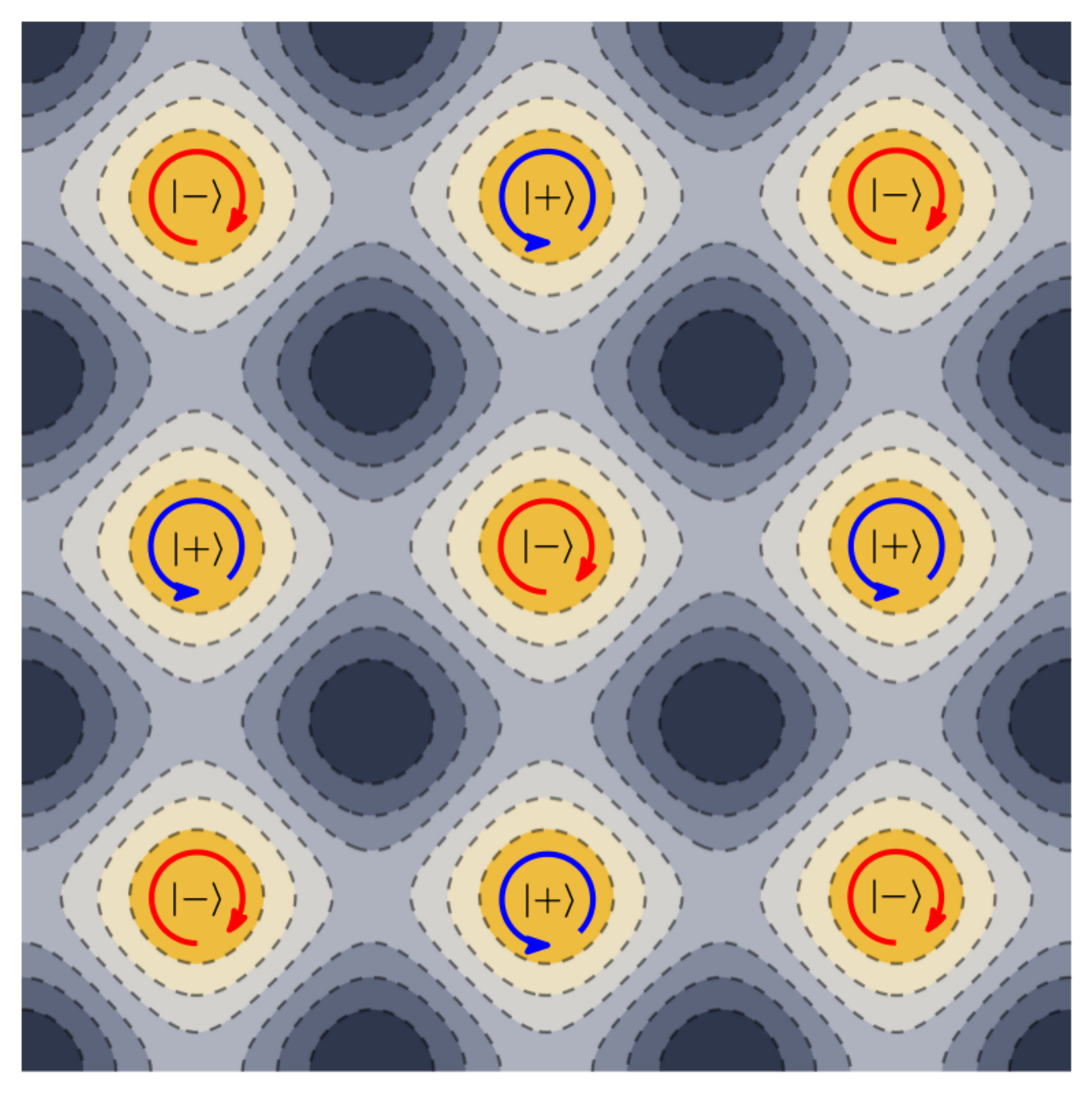}
\end{center}
\caption{Diagram of final state of each atom in the lattice. Each site contains one atom in state $\ket{\pm}$ with angular momentum  $\approx \pm \hbar$. \label{fig_state_diagram}}  
\end{figure}

The goal of this paper is to further develop the idea of shaking an optical lattice in order to create exotic states. This will be done by combining lattice shaking with techniques known as
``Shortcuts to Adiabaticity'' \cite{chen_2010}. 
In general, performing fast and stable state preparation of quantum systems is very demanding.
Adiabatic techniques are a common choice but have the drawback of needing extremely long times \cite{bergmann_1998}.
This has motivated the development of shortcuts to adiabaticity, which are protocols which reach fidelities of adiabatic processes in significantly shorter times. For a review of these see \cite{sta_review, sta_review_2}.
An important advantage of these methods is that they possess a certain freedom to optimise against noise, systematic error or unwanted transitions to higher levels \cite{ruschhaupt_2012, daems_2013, lu_2013, kiely_2014, lu_2014}.
In the following, we will show that combining optical lattice shaking with shortcut techniques can lead to schemes that are experimentally feasible (only requiring control over the relative phase and the polarisation of the lasers) and still have the freedom to be further optimised against the most relevant experimental noise sources.
In particular, we will choose a staggered order angular
momentum state as our target state, which has a lot of physically interesting properties \cite{isacsson_2005, liu_2006, hebert_2013,collin_2010}.
This non-trivial state has an anti--ferromagnetic type ordering, which consist of each potential well
being occupied by a single atom, carrying alternating angular momentum $\approx \pm \hbar$ (see figure \ref{fig_state_diagram}).
%
We will propose a method which, starting from a Mott--insulator state, prepares such an anti--ferromagnetic type ordering by shaking the lattice.  The state we create can be seen as a stepping stone towards more complex higher band states and the method we present is readily extendible to generate other states. It should be noted that shortcuts have been suggested previously for the creation of angular momentum in ultracold atom systems ~\cite{STA_ang1,STA_ang2}.

The remainder of this paper is structured as follows. In the subsequent
section, we outline our model for the shaken optical lattice. In Section
\ref{LR_review}, we review the method of Lewis--Riesenfeld invariants. In
Section \ref{schemes}, we outline the different schemes used in order to
prepare the angular momentum state. In Section \ref{Numerics}, we perform
numerical simulation of the full Schr\"{o}dinger equation for a single atom in one site of
an optical lattice in order to verify our assumptions. In Section \ref{exp_sec}, we remark on some experimental consideration. Finally in Section \ref{discuss}, we discuss our results.

\section{Model \label{model}}

\subsection{Optical lattice\label{optical}}
We consider a two--dimensional optical lattice (in the $x$--$y$ plane) generated by two pairs of counter--propagating laser beams.
We assume a strong confinement in the $z$ direction such that only dynamics in the $x$--$y$ plane are relevant.
We also assume that the atoms are in the Mott insulator regime i.e. each site is occupied by a single atom which is effectively independent of all the others.
One can enter such a regime by having a large lattice depth so that tunnelling rates are small.
While this means it is sufficient to consider each atom separately in the following, it is important to note that the all operations presented here are global and will affect all the atoms/sites simultaneously.

The complex amplitude of the electric field of the laser beams generating the two--dimensional optical lattice is
\begin{eqnarray}
\vec{\mathcal{E}}(x,y,t)=\vec{\mathcal{E}}_{0} \sin\left \{ k\left[x-r_{x}(t)\right]\right \}+ i \vec{\mathcal{E}}_{0} e^{-i \rho(t)} \sin\left \{ k\left[y-r_{y}(t)\right]\right \} ,
\end{eqnarray}
where $r_{x}(t)$ and $r_{y}(t)$ define the position of the minimum of the central trap, and can be controlled by a time-dependent phase difference between the pair of laser beams in each direction.
When these are modulated periodically, it results in a shaking of the lattice.
We will see below that this shaking alone is insufficient to create the desired quantum state.
Therefore, we assume in addition that the polarisation vectors in the two directions have an equal amplitude $\vec{\mathcal{E}}_{0}$, but with a slowly varying relative phase $\rho(t)$.

The potential felt by an atom in the two--dimensional optical lattice is given by \cite{lewenstein_2012}
\begin{eqnarray}
V(x,y)=\frac{1}{4 \hbar \Delta}\left|\vec\mu \cdot \vec{\mathcal{E}}^{*}(x,y,t)\right|^{2} ,
\end{eqnarray}
where $\vec\mu$ is the transition dipole moment of the atom and $\Delta$ (assumed to be large) is the detuning of the laser with respect to the atomic transition frequency.
Defining the lattice depth as
\begin{eqnarray}
V_0= \frac{1}{ 4 \hbar \Delta } \left|\vec\mu \cdot \vec{\mathcal{E}}_{0}^{*}\right|^{2}  ,
\end{eqnarray}
the potential can be written as
\begin{eqnarray}
V(x,y)=&& V_0 \sin^{2}\left\{k\left[x-r_{x}(t)\right]\right\}+V_0\sin^{2}\left\{k\left[y-r_{y}(t)\right]\right\} \nonumber \\
&&+V_\rho(t)\sin\left\{k\left[x-r_{x}(t)\right]\right\}\sin\left\{k\left[y-r_{y}(t)\right]\right\} ,
\end{eqnarray}
where $V_\rho(t)=2V_0\sin\left[\rho(t)\right]$ is the amplitude of the interference term, restricted to the interval $\left[-2V_0,2V_0\right]$.
Without any loss of generality, we assume that the laser is blue detuned ($\Delta >0$) so that $V_0$ is positive.

We now change from the lab frame to the lattice frame (see \ref{app} for details), where the Hamiltonian takes the form
\begin{eqnarray}
H_\textrm{lattice}(t) = H_{0}+H_{1}(t),\label{H_lattice_frame}\\
H_{0} = -\frac{\hbar^{2}}{2m}\nabla^{2}+V_0\sin^{2}(k x)+V_0\sin^{2}(k y),\\
H_{1}(t) = m \ddot{r}_{x}(t)x+m \ddot{r}_{y}(t)y+V_\rho(t)\sin(k x)\sin(k y). \label{H_1}
\end{eqnarray}
It is worth noting at this point that without the $V_\rho$ term, the Hamiltonian would be separable in $x$ and $y$ and therefore unable to create entanglement between the $x$ and $y$ degrees of freedom, which is necessary for angular momentum states.
We will assume the shaking of the lattice to be of the form
\begin{eqnarray}
r_{x}(t)&=&-g_{x}(t)\cos(\omega_{x}t) ,\nonumber \\
r_{y}(t)&=&g_{y}(t)\sin(\omega_{y}t) ,
\end{eqnarray}
where $g_{x,y}(t)$ are the time--dependent amplitudes and $\omega_{x,y}$ are the frequencies.
By assuming that $g_{x,y}(t)$ vary slowly with time,
$H_{1}(t)$ simplifies to
\begin{eqnarray}
H_{1}(t)&=&f_x(t)\,x + f_y (t)\,y +V_\rho(t)\sin(k x)\sin(k y) ,
\end{eqnarray}
where
\begin{eqnarray}
f_{x}(t)=m \omega_{x}^{2}g_{x}(t)\cos(\omega_{x}t), \\
f_{y}(t)=-m \omega_{y}^{2}g_{y}(t)\sin(\omega_{y}t).
\end{eqnarray}
In this case the shaking in the $y$ direction is $\pi/2$ out of phase with the shaking in $x$ direction.

\subsection{Four--level approximation \label{four_level_approx}}

\begin{figure}[t]
\begin{center}
\includegraphics[angle=0,width=0.45\linewidth]{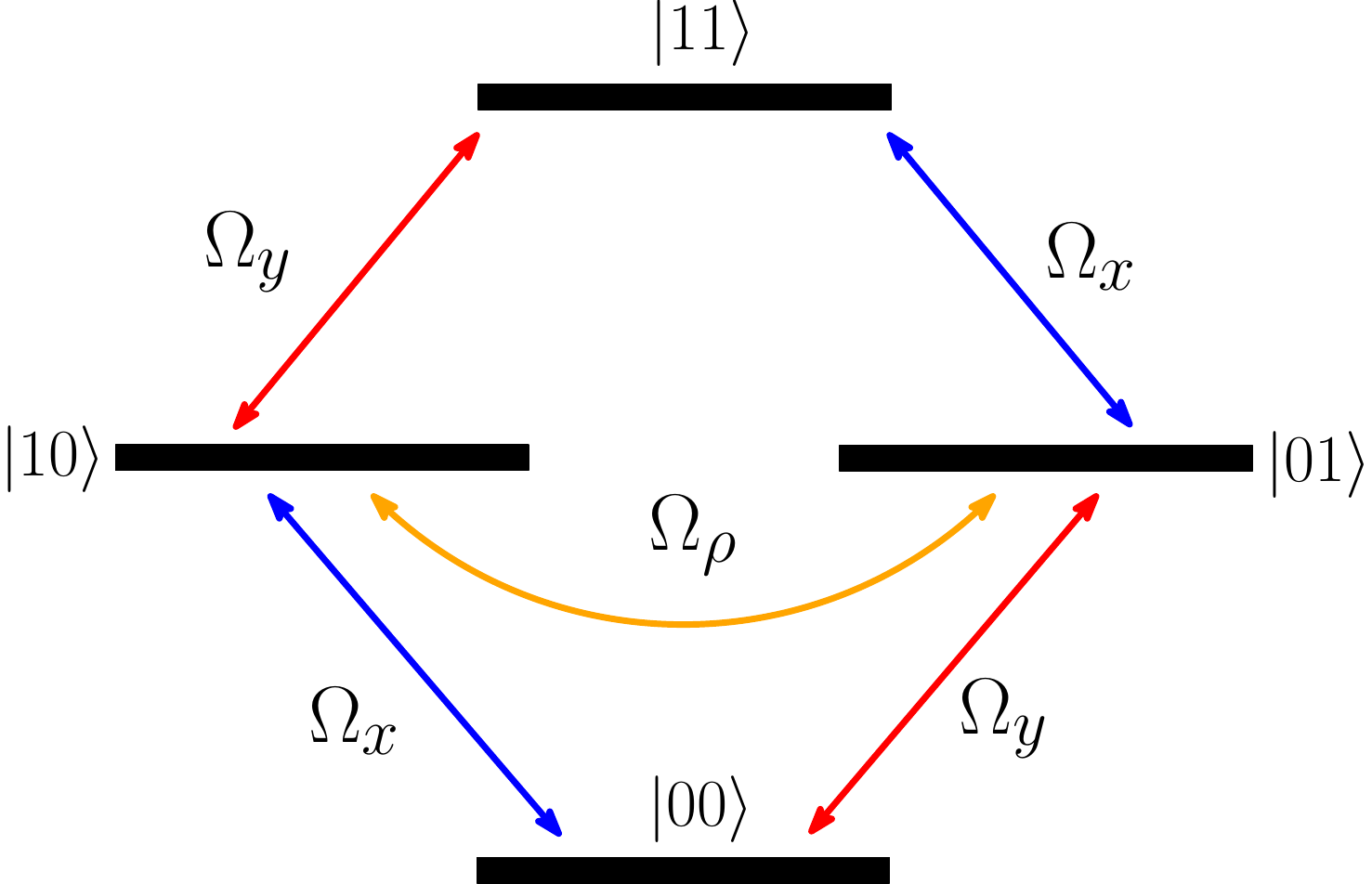}

\end{center}
\caption{Energy level diagram for the four chosen energy eigenstates of $H_{0}$ and the various couplings between them.
 \label{fig_level}}  
\end{figure}

Our aim is to derive the control schemes, i.e., the time dependence of the functions $r_x(t)$, $r_y(t)$ and $V_\rho(t)$, which will lead to a desired final state.
To do this we will now derive a simplified model of the system by concentrating on a single atom in a single well of the lattice defined by
$- \ell\le x \le \ell$ and $-\ell\le y \le \ell$, 
where $2\ell = \pi/k$ is the lattice constant. The situation where the neighbouring lattice potential wells can be neglected is very well realised in the Mott insulator regime.

Furthermore, we assume that the dynamics can be effectively described by a four--level approximation, considering only the four most relevant eigenstates of $H_{0}$ localised in the central site, 
 $\left\{\ket{00},\ket{10},\ket{01},\ket{11}\right\}$
(see figure \ref{fig_level}). The validity of this and all subsequent approximations will be checked later by comparing with the numerical integration of the full Schr\"odinger equation.
In coordinate representation, these basis states are given by
\begin{eqnarray}
\braket{\vec{r}}{ij}=\Gamma_{i}(x)\Gamma_{j}(y) ,
\end{eqnarray}
where $\Gamma_{0}(x)$ and $\Gamma_{1}(x)$ are, respectively, the localised ground and first excited states of a one--dimensional unperturbed optical lattice site.
Note that this is only possible because $H_{0}$ is separable in $x$ and $y$.
Their respective energies are $E_{ij}=\hbar \omega_{ij}$,
where $E_{00}<E_{01}=E_{10}<E_{11}$.

Let us now define a unitary transformation of the form
\begin{eqnarray}
U(t)= && e^{-i( \omega_{10}+\omega_{x})t}\ket{00}\bra{00}+e^{-i( \omega_{10}+\omega_{x}-\omega_{y})t}\ket{01}\bra{01}
\nonumber \\ && +e^{-i \omega_{10}t}\ket{10}\bra{10}+e^{-i \omega_{11}t}\ket{11}\bra{11} ,
\end{eqnarray}
under which the Hamiltonian changes as
\begin{eqnarray}
H &\longrightarrow & U^{\dagger}HU-i \hbar U^{\dagger}\dot{U} =
  U^{\dagger}H_{0}U -i \hbar U^{\dagger}\dot{U}  +U^{\dagger}H_{1}(t)U = H_{4L} .
\end{eqnarray}
The first part of this is
\begin{eqnarray}
U^{\dagger}H_{0}U-i \hbar U^{\dagger}\dot{U}=\hbar(\omega_{00}-\omega_{10}-\omega_{x})\ket{00}\bra{00}
+\hbar(\omega_{y}-\omega_{x})\ket{01}\bra{01},
\end{eqnarray}
and the second part simplifies to
\begin{eqnarray}
U^{\dagger}H_{1}(t)U = && e^{-i \omega_{x} t}\gamma_{1} f_{x}(t) \ket{10}\bra{00}+V_\rho(t) \gamma_{2} e^{-i(\omega_{x}-\omega_{y})t}\ket{10}\bra{01}\nonumber \\
&&+e^{i(\omega_{x}-\omega_{y}-\omega_{d})t}\gamma_{1} f_{x}(t)\ket{01}\bra{11}+e^{-i\omega_{d}t}\gamma_{1} f_{y}(t) \ket{10}\bra{11}
\nonumber \\
&&+e^{i\omega_{y}t}\gamma_{1} f_{y}(t)\ket{00}\bra{01}+V_\rho(t) \gamma_{2} e^{i(\omega_{x}-\omega_{d})t}\ket{00}\bra{11}+\textrm{h.c.} ,
\end{eqnarray}
where we have defined
\begin{eqnarray}
\gamma_{1} &=&\int_{-\ell}^{\ell} \Gamma_{0}(x)x\Gamma_{1}(x)dx , \\
\gamma_{2} &=&\left[\int_{-\ell}^{\ell} \Gamma_{0}(x)\sin(k x)\Gamma_{1}(x)dx\right]^{2} , \\
\omega_{d} &=& \omega_{10}-\omega_{00} .
\end{eqnarray}
Note that the symmetry of the unperturbed lattice gives $\omega_{11}=2\omega_{10}-\omega_{00}$.

We now assume that the shaking of the lattice in both directions is done on resonance, i.e., $\omega_{x}=\omega_{y}=-\omega_{d}$.
This allows to write the four--level Hamiltonian as
\begin{eqnarray}
H_{4L}(t)= \frac{\hbar}{2}\biggl[&& \Omega_{x}(t)\left(1+e^{2i\omega_{d}t}\right)\ket{10}\bra{00}
+\Omega_{x}(t)\left(1+e^{-2i\omega_{d}t}\right)\ket{01}\bra{11}\nonumber \\
&&  -i \Omega_{y}(t)\left(1-e^{-2i\omega_{d}t}\right)\ket{10}\bra{11}
-i \Omega_{y}(t)\left(1-e^{-2i\omega_{d}t}\right)\ket{00}\bra{01}\nonumber \\
&&  +\Omega_{\rho}(t)\ket{10}\bra{01}+\Omega_{\rho}(t) e^{-2 i \omega_{d}t}\ket{00}\bra{11}+\textrm{h.c.}
\biggr] ,
\end{eqnarray}
with the couplings
\begin{eqnarray}
\Omega_{x,y}(t)&=&m \omega_{d}^{2}\gamma_{1} g_{x,y}(t)/\hbar, \nonumber \\
\Omega_\rho(t)&=&2 V_\rho(t)\gamma_{2} /\hbar .
\end{eqnarray}
By making a rotating wave approximation,
where the terms containing $e^{\pm 2 i \omega_{d}t}$ average to 0,
we arrive at our final four--level Hamiltonian (see figure \ref{fig_level})
\begin{eqnarray}
H_{4L}(t)=\frac{\hbar}{2}\left(\begin{array}{cccc}
0 & \Omega_{x} & \Omega_\rho & -i\Omega_{y}\\
\Omega_{x} & 0 & -i\Omega_{y} & 0\\
\Omega_\rho & i\Omega_{y} & 0 & \Omega_{x}\\
i\Omega_{y} & 0 & \Omega_{x} & 0
\end{array}\right),\label{H_4L}
\end{eqnarray}
where we have used the following representation of the states
\begin{eqnarray}
\ket{10} = \left(\begin{array}{c}
1\\
0\\
0\\
0
\end{array}\right),\enspace
\ket{00} = \left(\begin{array}{c}
0\\
1\\
0\\
0
\end{array}\right),\enspace
\ket{01} = \left(\begin{array}{c}
0\\
0\\
1\\
0
\end{array}\right),\enspace
\ket{11} = \left(\begin{array}{c}
0\\
0\\
0\\
1
\end{array}\right).
\end{eqnarray}
It is important to note that state $\ket{11}$ can not be neglected and should be included in the approximation, as it is resonantly coupled to $\ket{01}$ and $\ket{10}$.

\subsection{Initial and target states}

Our goal is to perform a state transfer from the ground state $\ket{00}$ to an angular momentum state of the form
\begin{eqnarray}
\ket{\pm}=\frac{1}{\sqrt{2}}\left(\ket{10}\pm i\ket{01}\right) .
\end{eqnarray}
If the harmonic approximation holds, $\ket{\pm}$ are eigenvectors of the $z$ component of the angular momentum operator $L_{z}$ with eigenvalues $\pm \hbar$.

One can see that the interference term in \eqref{H_1}, which includes $V_\rho$, alternates sign at each lattice site in a checkerboard pattern. In the case where $\Omega_{y}=0$, this can be seen as a change of basis $\ket{01} \rightarrow -\ket{01}$ and $\ket{11} \rightarrow -\ket{11}$ and hence one obtains either $\ket{+}$ or $\ket{-}$ in alternating sites, leading to the pattern in Fig.~\ref{fig_state_diagram}. For our schemes we will assume that $\Omega_{y}=0$, although more general schemes might be derived in a similar way.

In the following, we will use the technique of Lewis--Riesenfeld invariants to derive shortcut schemes to implement the state transfer $\ket{00} \rightarrow \ket{-}$. An advantage of this method is that one still has a certain freedom to optimise the stability of the schemes against the most relevant error sources in a specific setting \cite{ruschhaupt_2012, daems_2013, lu_2013, kiely_2014, lu_2014}.


\section{Lewis--Riesenfeld invariants for the four--level system\label{LR_review}}

One possible technique to derive shortcuts to adiabaticity is based on Lewis--Riesenfeld invariants \cite{LR69}.
A Lewis--Riesenfeld invariant for a Hamiltonian $H(t)$ is a Hermitian operator $I(t)$ which satisfies
\begin{equation}
\frac{\partial I}{\partial t}+\frac{i}{\hbar}\left[H,I\right]=0.
\label{eq:LR}
\end{equation}
Since $I(t)$ is a constant of motion it can be shown that it has time--independent eigenvalues and that a particular solution of the Schr\"{o}dinger equation,
\begin{equation}
i\hbar \frac{\partial}{\partial t}\left|\psi_n(t)\right\rangle = H(t) \left|\psi_n(t)\right\rangle,
\end{equation}
 can be written as
\begin{equation}
\left|\psi_{n}(t)\right\rangle=e^{i \beta_{n}(t)} \left|\phi_{n}(t)\right\rangle.
\end{equation}
Here $\left|\phi_{n}(t)\right\rangle$ is an instantaneous eigenstate of $I(t)$ and
\begin{equation}
\beta_{n}(t)=\frac{1}{\hbar} \int_{0}^t \left\langle
\phi_{n}(s)\right.\left|\left[i\hbar\partial_{s}-H(s)\right]\right|\left.\phi_{n}(s)\right\rangle ds\,
\end{equation}
is the Lewis--Riesenfeld phase.
Hence a general solution to the Schr\"{o}dinger equation can be written as
\begin{equation}
\left|\psi(t)\right\rangle=\sum_{n} c_{n} \left|\psi_{n}(t)\right\rangle
\end{equation}
where the $c_{n}$ are independent of time.

The idea behind inverse engineering is that instead of following the instantaneous eigenstate of the Hamiltonian (as in the adiabatic case), one follows the instantaneous eigenstate of the invariant (up to the Lewis--Riesenfeld phase).
Demanding that the invariant and the Hamiltonian commute at the start and the end of the process i.e., $\left[I(0),H(0)\right]=\left[I(T),H(T)\right]=0$, one ensures that the eigenstates of the invariant and the Hamiltonian coincide at initial and final times.
This leaves the freedom to choose how the state evolves in the intermediate time and then use \eqref{eq:LR} to determine how the Hamiltonian should vary with time to ensure such a state evolution.

In the following we will derive the invariant for the Hamiltonian in \eqref{H_4L} with $\Omega_{y}=0$. For a more detailed review of Lewis--Riesenfeld invariants for four level systems see \cite{four_invariants}. 
Following the general method proposed in \cite{torrontegui_2014,martinez_2014},
we start with a closed Lie algebra
$\{ G_{1},G_{2},G_{3},G_{4} \}$ of Hermitian operators
\begin{eqnarray}
G_{1} &= \left(\begin{array}{cccc}
0 & 1 & 0 & 0\\
1 & 0 & 0 & 0\\
0 & 0 & 0 & 1\\
0 & 0 & 1 & 0
\end{array}\right)\,,
G_{2} = \left(\begin{array}{cccc}
0 & 0 & 1 & 0\\
0 & 0 & 0 & 0\\
1 & 0 & 0 & 0\\
0 & 0 & 0 & 0
\end{array}\right)\,,\nonumber\\
G_{3} &= \left(\begin{array}{cccc}
0 & 0 & 0 & i\\
0 & 0 & -i & 0\\
0 & i & 0 & 0\\
-i & 0 & 0 & 0
\end{array}\right)\,,
G_{4} = \left(\begin{array}{cccc}
0 & 0 & 0 & 0\\
0 & 0 & 0 & 1\\
0 & 0 & 0 & 0\\
0 & 1 & 0 & 0
\end{array}\right)\,.
\end{eqnarray}

The $4$--level Hamiltonian and the associated Lewis--Riesenfeld invariant can then be written as a linear combination of these operators
\begin{eqnarray}
H\left(t\right)=\frac{\hbar}{2}\Omega_{x}\left(t\right)G_{1}+\frac{\hbar}{2}\Omega_\rho\left(t\right)G_{2} , \\
I\left(t\right)=\sum^{4}_{i=1} \alpha_{i}\left(t\right) G_{i} ,
\end{eqnarray}
where $\alpha_{i}(t) \in\mathbb{R}$.
Inserting this into \eqref{eq:LR}, we get that the coupling strengths are given by
\begin{eqnarray}
\Omega_{x}\left(t\right)= -\frac{\dot{\alpha}_{2}(t)}{\alpha_{3}(t)},
\label{omega_x}\\
\Omega_\rho\left(t\right)= \frac{2 \dot{\alpha}_{1}(t)}{\alpha_{3}(t)},
\label{omega_rho}
\end{eqnarray}
and that
\begin{eqnarray}
\alpha_{3}(t)=\xi \sqrt{2 C_{2}-[\alpha_{1}^{2}(t)+\alpha_{2}^{2}(t)]+ C_{1} \alpha_{2}(t) }, \\
\alpha_{4}(t)= C_{1}-\alpha_{2}(t),
\end{eqnarray}
where $C_{1,2} \in \mathbb{R}$ are constants, $\xi=\pm 1$ and $\alpha_{1}(t), \alpha_{2}(t)$ are still arbitrary functions.

In order to be useful it is important to know the eigenvalues $\kappa_{i}$ and eigenvectors $\ket{\phi_{i}(t)}$
of the invariant, i.e. $I(t)=\sum_{i=1}^{4}\kappa_{i}\ketbra{\phi_{i}(t)}{\phi_{i}(t)}$. 
We get that the eigenvalues are
\begin{eqnarray}
\kappa_{1}&=&\frac{1}{2}\left(-C_{1}-Q\right), \,\kappa_{2}=\frac{1}{2}\left(C_{1}-Q\right),\nonumber\\
\kappa_{3}&=&\frac{1}{2}\left(-C_{1}+Q\right), \,\kappa_{4}=\frac{1}{2}\left(C_{1}+Q\right),
\end{eqnarray}
where $Q=\sqrt{C_{1}^{2}+8 C_{2}}$.
The corresponding eigenvectors are
\begin{eqnarray}
\left|\phi_{1}\left(t\right)\right\rangle =\left(\begin{array}{c}
-B_{+}D_{-}\\
-\frac{1}{2B_{+}}\\
B_{+}D_{-}\\
\frac{1}{2B_{+}}
\end{array}\right)\,,
\left|\phi_{2}\left(t\right)\right\rangle =\left(\begin{array}{c}
-B_{-}D_{+}\\
\frac{1}{2B_{-}}\\
-B_{-}D_{+}\\
\frac{1}{2B_{-}}
\end{array}\right)\,,\\
\left|\phi_{3}\left(t\right)\right\rangle =\left(\begin{array}{c}
B_{-}D_{-}\\
-\frac{1}{2B_{-}}\\
-B_{-}D_{-}\\
\frac{1}{2B_{-}}
\end{array}\right)\,,
\left|\phi_{4}\left(t\right)\right\rangle =\left(\begin{array}{c}
B_{+}D_{+}\\
\frac{1}{2B_{+}}\\
B_{+}D_{+}\\
\frac{1}{2B_{+}}
\end{array}\right)\,,
\end{eqnarray}
where we have defined
\begin{eqnarray}
B_{\pm}(t)&=& \sqrt{\frac{Q}{\pm C_{1}+Q \mp 2 \alpha_{2}}} , \\
D_{\pm}(t)&=& \frac{i}{Q}\left[\frac{2 C_{2}+(C_{1}-\alpha_{2})\alpha_{2}}{\pm i \alpha_{1}+\xi \sqrt{2 C_{2}+(C_{1}-\alpha_{2})\alpha_{2}-\alpha_{1}^{2}}}\right] .
\end{eqnarray}
Note that $Q,B_{\pm}\in \mathbb{R}$ and $D_{+}^{*}=-D_{-}$. We also assume a nonzero $Q$ so that none of the above quantities diverge.

Finally, the Lewis-Riesenfeld phases~\cite{LR69} are given by 
\begin{eqnarray}
\beta_{1}(t)&=-\chi_{+}(t),\, &\beta_{2}(t)=\chi_{-}(t),\nonumber\\
\beta_{3}(t)&= -\chi_{-}(t),\, &\beta_{4}(t) = \chi_{+}(t) \, ,
\end{eqnarray}
where we have defined
\begin{eqnarray}
\chi_{\pm}(t)= \int_{0}^{t} \frac{2\alpha_{1}\left[C_{1}^{2}+4C_{2}\pm C_{1}Q \mp 2\left(\pm C_{1}+Q\right)\alpha_{2}+2\alpha_{2}^{2}\right]\dot{\alpha}_{2}}{\left(C_{1}\pm Q-2 \alpha_{2}\right)^{3}\xi \left[2 C_{2}+\left(C_{1}-\alpha_{2}\right)\alpha_{2}-\alpha_{1}^{2}\right]^{\frac{1}{2}}} ds .
\end{eqnarray}


\section{Shaking schemes for preparing an angular momentum state\label{schemes}}

\begin{figure}[t]
\begin{center}
\includegraphics[angle=0,width=0.45\linewidth]{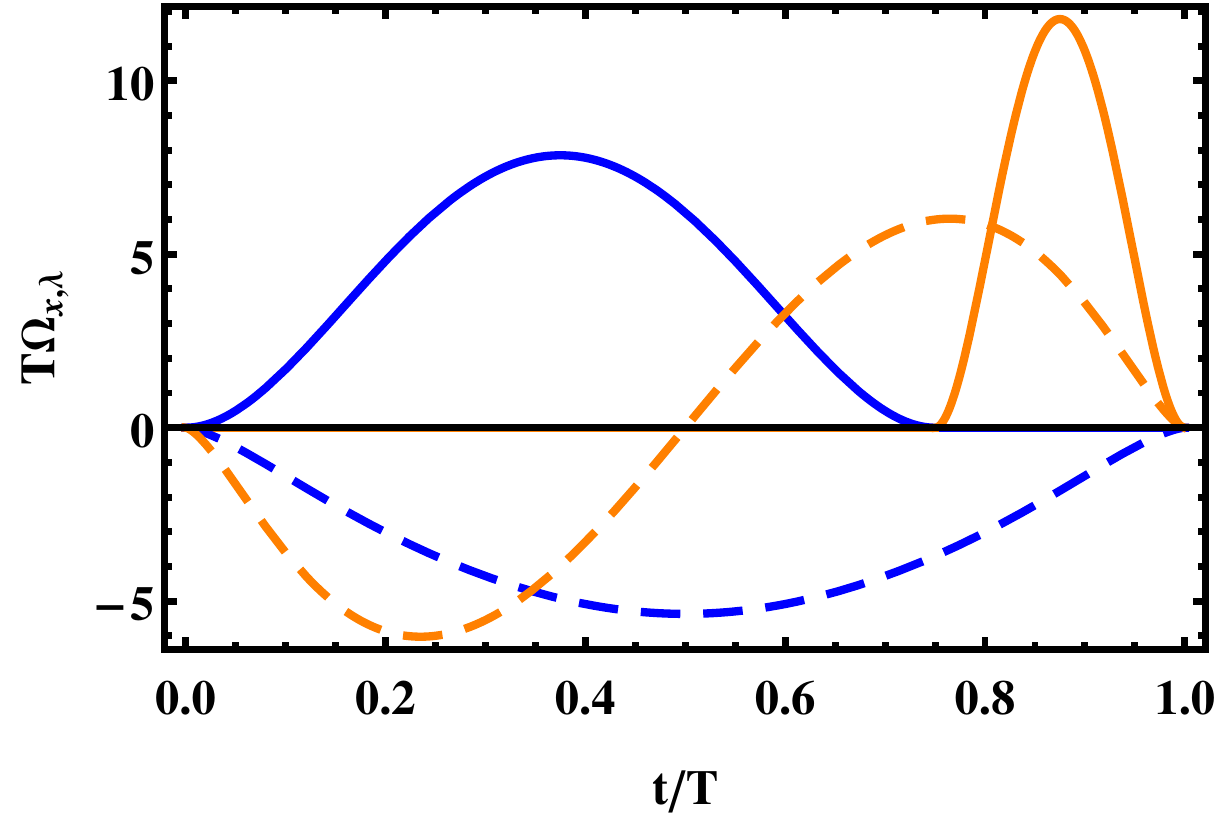}

\end{center}
\caption{Coupling strengths against time for the two different schemes.  Polynomial scheme: $\Omega_{x}$ (blue, dashed line) and $\Omega_\rho$ (orange, dashed line). Piecewise scheme ($t_S=0.75T$): $\Omega_{x}$ (blue, solid line) and $\Omega_\rho$ (orange, solid line). \label{fig_schemes}}  
\end{figure}

In this section, we present two schemes which allow us to prepare our target state.
In order to design the scheme we start by constructing a solution to the Schr\"odinger equation as a linear combination of two of the eigenvectors of the invariant
\begin{eqnarray}
\left|\psi\left(t\right)\right\rangle &=&\frac{1}{\sqrt{2}}\left[-\left|\phi_{1}\left(t\right)\right\rangle e^{i \beta_{1}(t)}+\left|\phi_{4}\left(t\right)\right\rangle e^{i \beta_{4}(t)}\right] \nonumber \\
&=&\frac{1}{\sqrt{2}}\left[-\left|\phi_{1}\left(t\right)\right\rangle e^{-i \beta_{4}(t)}+\left|\phi_{4}\left(t\right)\right\rangle e^{i \beta_{4}(t)}\right].
\end{eqnarray}
The initial and final state of the system are fixed as
\begin{eqnarray}
\ket{\psi(0)}=\ket{00} , \\
\ket{\psi(T)}=\ket{-},
\end{eqnarray}
which leads to the boundary conditions
\begin{eqnarray}
\alpha_{1}(0) = 0 ,\, \alpha_{2}(0) = (C_{1}-Q)/2, \\
\alpha_{1}(T) = 0 ,\, \alpha_{2}(T)=(C_{1}+Q)/2 ,\,
\beta_{4}(T)= 0,
\end{eqnarray}
in the limits $t\to 0$ and $t\to T$.

We also demand that $\Omega_{x}$, $\Omega_\rho$ and their respective derivative are zero at the start and the end of the process.
This requires that all the derivatives of $\alpha_{1}(t)$ and $\alpha_{2}(t)$ up to fourth order are zero at $t=0$ and $t=T$, which gives 10 constraints to be fulfilled by $\alpha_{1}(t)$ and also 10 constraints for $\alpha_{2}(t)$.

\subsection{Polynomial scheme}

A convenient choice of ansatz for $\alpha_{1}(t)$ and $\alpha_{2}(t)$ which fulfills all the constraints is given by polynomials of the form
\begin{eqnarray}
\alpha_{1}(sT)= 1024W (-s^{10}+5s^9-10s^8+10s^7-5s^6+s^5), \\
\alpha_{2}(sT)=\frac{1}{2}\left(C_{1}-Q\right)+70Qs^9-315Qs^8
 +540Qs^7-420Qs^6+126Qs^5, \nonumber
\end{eqnarray}
where $s=t/T$. To avoid the trivial solution $\alpha_1(sT)=0$ we also demand $\alpha_1(T/2)=W\neq 0$.
We are now allowed to arbitrarily pick $C_{1}=10$ and $C_{2}=11$ so that $Q \neq 0$ and $\alpha_{3}(t)$ is real for all times.
We also set $\xi=+1$ and then numerically calculate $W$ ($\approx -2.74$) 
so that $\beta_{4}(T)=0$.
The coupling strengths $\Omega_{x}(t)$ and $\Omega_\rho(t)$ can be calulated from Eqs. \eqref{omega_x} and \eqref{omega_rho}, and are shown in figure \ref{fig_schemes} (dashed lines).

Let us underline again that this is just one possible choice for the auxiliary functions $\alpha_1(t)$ and $\alpha_2(t)$ (and the constants $C_1$ and $C_2$). The advantage of this inverse--engineering ansatz is that it provides a lot of freedom in choosing these functions which can be used for further optimisations \cite{ruschhaupt_2012}.

\subsection{Piecewise scheme}
The second example we introduce to generate our target state is a simple piecewise scheme.
The idea is to first perform a $\pi$ pulse in $\Omega_{x}$ (of duration $t_S$) which transfers all the population from $\ket{00}$ to $\ket{10}$, followed by a $\pi/2$ pulse in $\Omega_\rho$ (of duration $T-t_S$) which leads to the superposition $\ket{-}$.
This method has the advantage that the state $\ket{11}$ is never populated, which reduces the chance of losing population to higher levels. The amplitudes of the couplings are determined by $t_S$ and are given by (see figure \ref{fig_schemes} (solid lines))
\begin{eqnarray}
\Omega_{x}(t)&=& \begin{cases} 
\frac{30 \pi  t^2 (t-t_S)^2}{t_S^5}  & 0\leq t \leq t_S , \\ 0 & t_S < t\leq T ,
   \end{cases}\nonumber \\
\Omega_\rho(t)&=&\begin{cases} 
      0 & 0\leq t < t_S  ,\\
      -\frac{15 \pi (t - T)^2 (t - t_S)^2}{(t_S - T )^5} & t_S\leq t \leq T .
\end{cases}
\label{omegapiece}
\end{eqnarray}
Since $\Omega_{x}$ and $\Omega_{\rho}$ are a $\pi$ pulse and $\pi/2$ pulse respectively, we have that $\int_{0}^{T}\Omega_{x}(t)dt=\pi$ and $\int_{0}^{T}\Omega_{\rho}(t)dt=\pi/2$.

This can be seen as a particular case of schemes derived using invariant--based inverse engineering. In this case $\alpha_{1}(t)$ and $\alpha_{2}(t)$ are given by
\begin{eqnarray}
\fl \alpha_{1}(t)&=& \begin{cases} 
\epsilon  & 0\leq t \leq t_S , \\ \epsilon \cos\left[\frac{1}{2}\int_{t_S}^{t} \Omega_{\rho}\left(t'\right) dt'\right] & t_S < t\leq T ,
   \end{cases}\nonumber \\   
\fl \alpha_{2}(t)&=&\begin{cases} 
      \frac{1}{2}\left\{C_{1}-\sqrt{C_{1}^{2}+8 C_{2}-4 \epsilon^{2}}\cos\left[\int_{0}^{t} \Omega_{x}\left(t'\right) dt'\right]\right\} & 0\leq t < t_S  ,\\
      \frac{1}{2}\left(C_{1}+\sqrt{C_{1}^{2}+8 C_{2}-4 \epsilon^{2}}\right) & t_S\leq t \leq T ,
\end{cases}
\label{alpha}
\end{eqnarray}
and $\xi=-1$. Inserting Eqs. \eqref{alpha} in Eqs. \eqref{omega_x} and \eqref{omega_rho} gives back Eqs. \eqref{omegapiece}.
The required boundary conditions of $\alpha_1$ and $\alpha_2$ are fulfilled in the limit $\epsilon \rightarrow 0^{+}$.

\section{Numerical simulations of the shaking schemes\label{Numerics}}

The presented schemes result in the desired state transfer exactly in the framework of the four--level Hamiltonian.
In order to check the validity of all the approximations we have made to reach this model,
we present below simulations of the full Schr\"{o}dinger equation with Hamiltonian \eqref{H_lattice_frame} in coordinate space for an atom initially in the ground state of a single lattice site.

The evolution is performed by means of the Fourier split--operator method \cite{split_op}, where the initial ground state is found by imaginary--time evolution.
In order to make all plots dimensionless we define $\omega=\sqrt{\frac{2V_0k^{2}}{m}}$, which is the frequency of the harmonic oscillator potential which approximates each well of the optical lattice.
Note that the previously defined $\omega_{d}= \omega_{10} - \omega_{00}$ converges to $\omega$ for increasing lattice depth $V_0$.
The rotating wave approximation and the slowly--varying shaking amplitude approximation can be combined in the condition $T \gg \omega_{d}^{-1} \approx \omega^{-1}$.

As we have assumed to be in the Mott--insulator regime, we restrict our simulations to the dynamics of an atom in a single well.
We have checked the validity of this approximation by simulating our schemes in a $3 \times 3$ lattice.
With the typical parameters used below, the shaking causes only about a $1\%$ leakage into the neighbouring traps.

\begin{figure}[t]
\begin{center}
(a)\includegraphics[angle=0,width=0.6\linewidth]{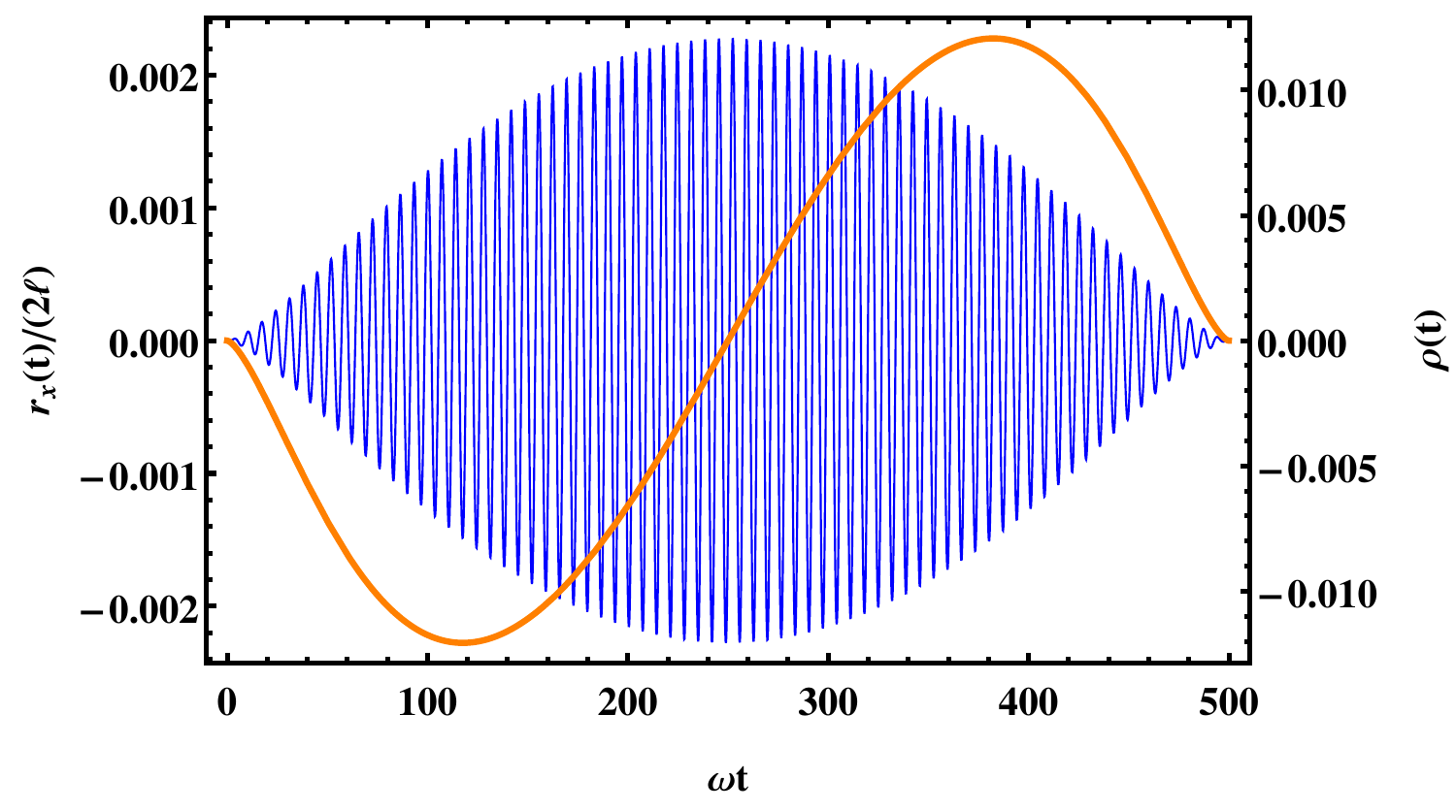}\\[1cm]

(b)\includegraphics[angle=0,width=0.6\linewidth]{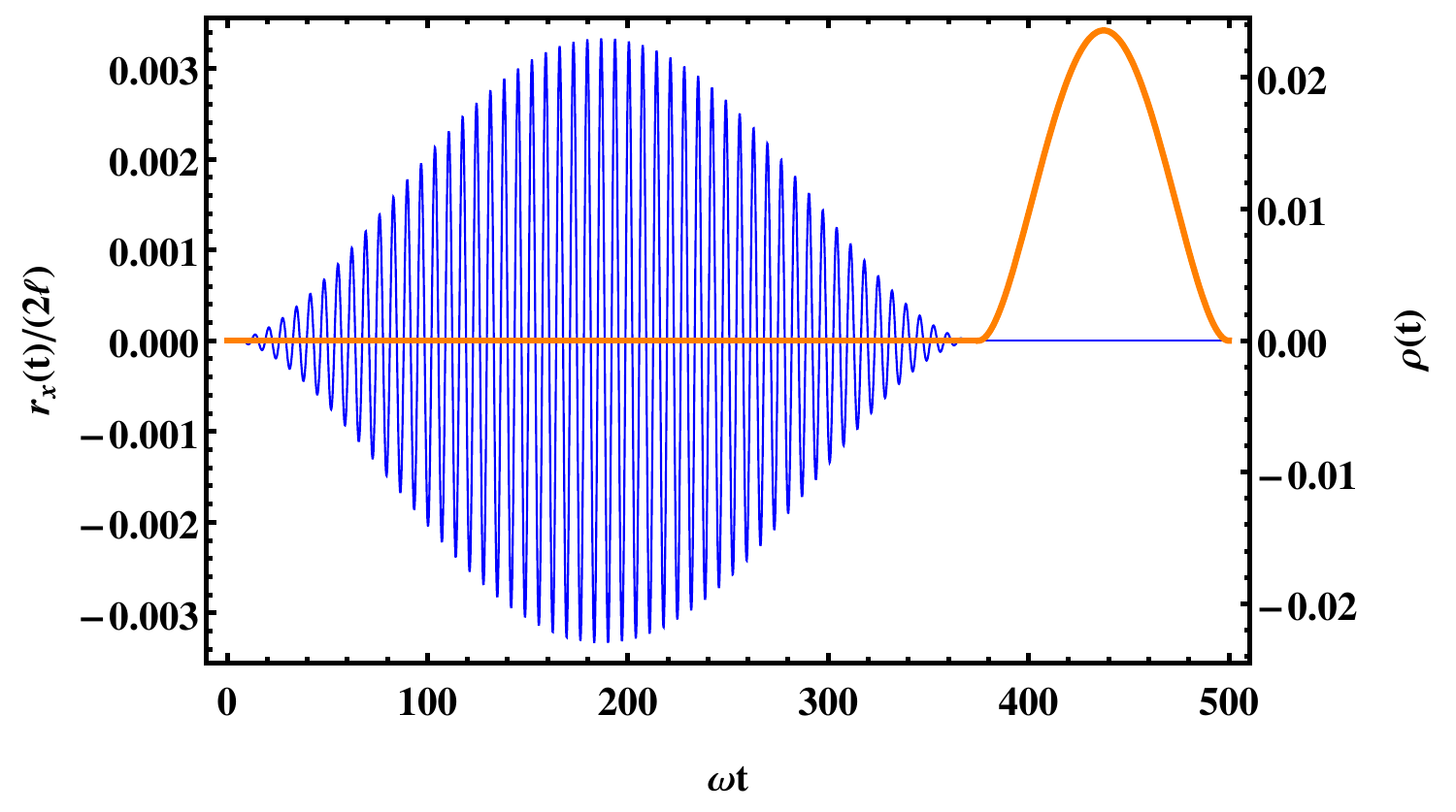}

\end{center}
\caption{Shaking function $r_x(t)$ with $\omega_{x}=-\omega_{d}$ (thin, blue line) and relative phase between the polarisation vectors $\rho(t)$ (thick, orange line) versus time for
(a) the polynomial scheme and (b) the piecewise scheme ($t_S=0.75T$). $V_0 = 3\hbar\omega$, $T=500 \omega^{-1}$ and $2 \ell=\pi/k$ is the lattice constant. \label{real_fkt}}  
\end{figure}

The control parameters in our system are the shaking function in the $x$ direction, $r_x(t)$ (as stated above, we keep $r_y(t) = 0$), and the relative phase between the polarisation vectors in the $x$ and $y$ directions, $\rho(t)$.
They relate to the couplings as
\begin{eqnarray}
r_x(t) = -\frac{\hbar}{m \omega_{d}^2 \gamma_{1}} \Omega_x (t) \cos\left(\omega_{x}t\right) , \\
\rho(t) = \arcsin\left(\frac{\hbar}{4V_0 \gamma_{2}} \Omega_\rho(t)\right) .
\end{eqnarray}
The resulting functions for both the polynomial process and the piecewise process  are shown in figure \ref{real_fkt}. One can see that the required amplitude of the shaking is only a small fraction of the lattice constant.

\begin{figure}[t]
\begin{center}
(a)\includegraphics[angle=0,width=0.43\linewidth]{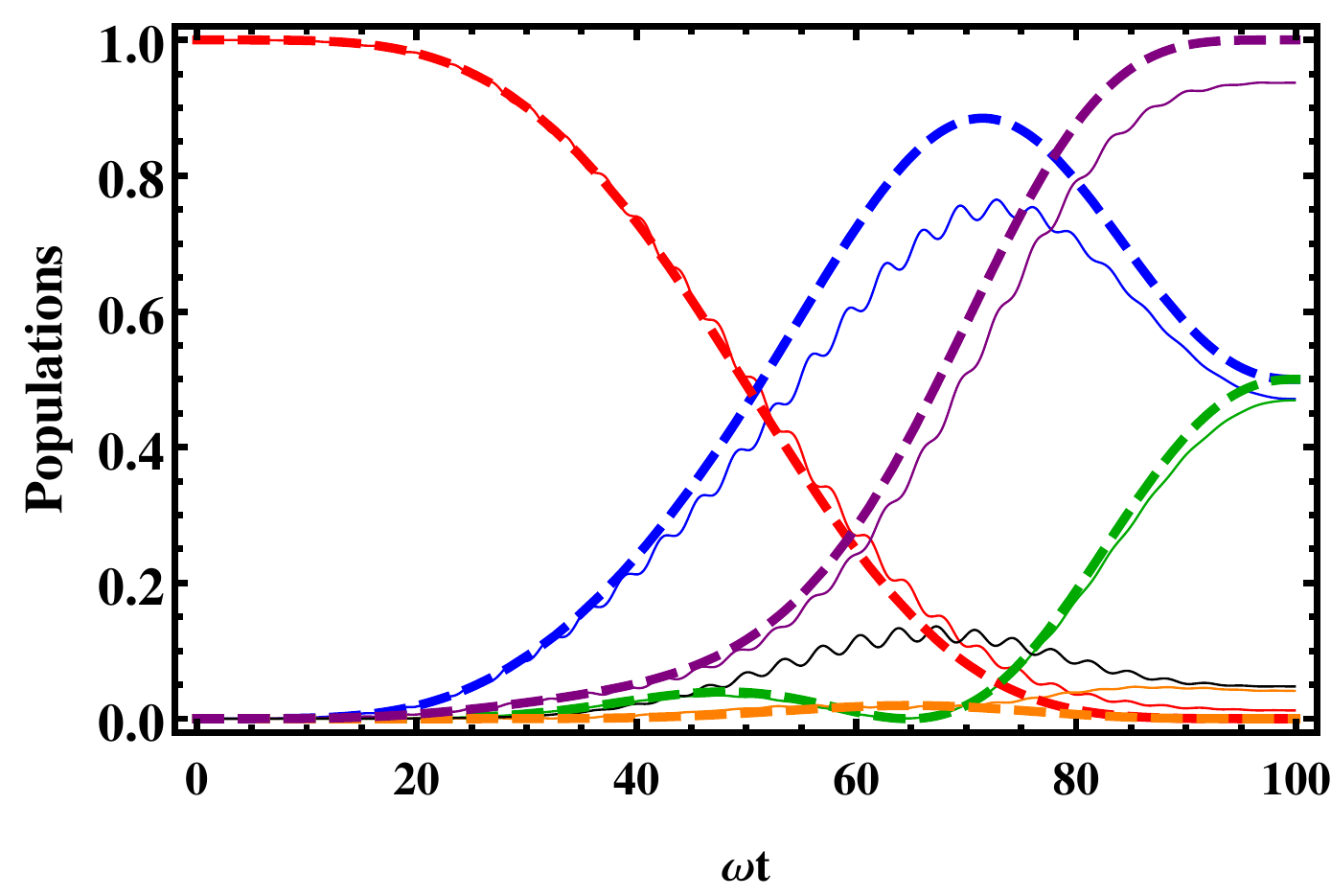}
\hspace{0.5cm}
(b)\includegraphics[angle=0,width=0.43\linewidth]{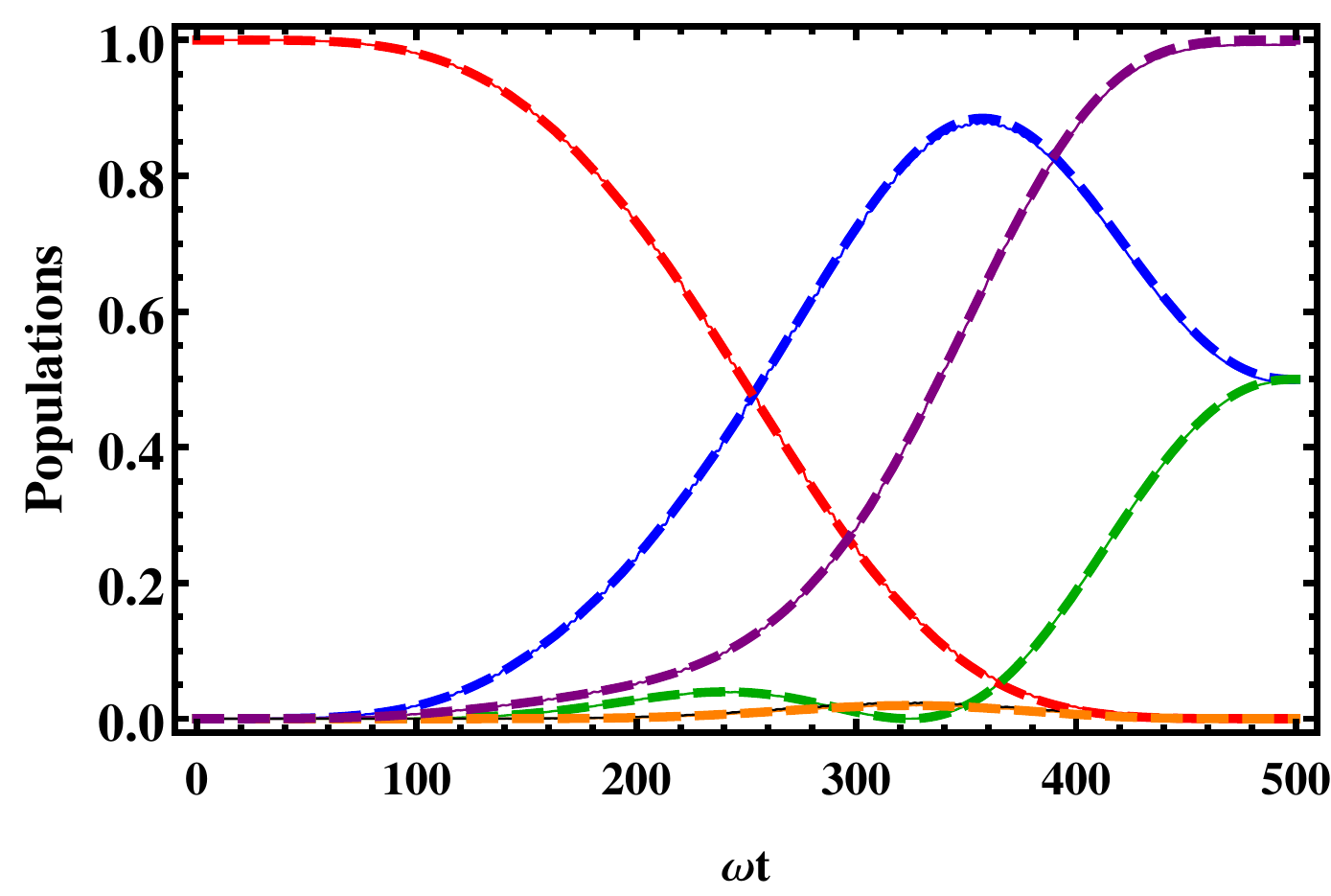}
\\[1cm]

(c)\includegraphics[angle=0,width=0.43\linewidth]{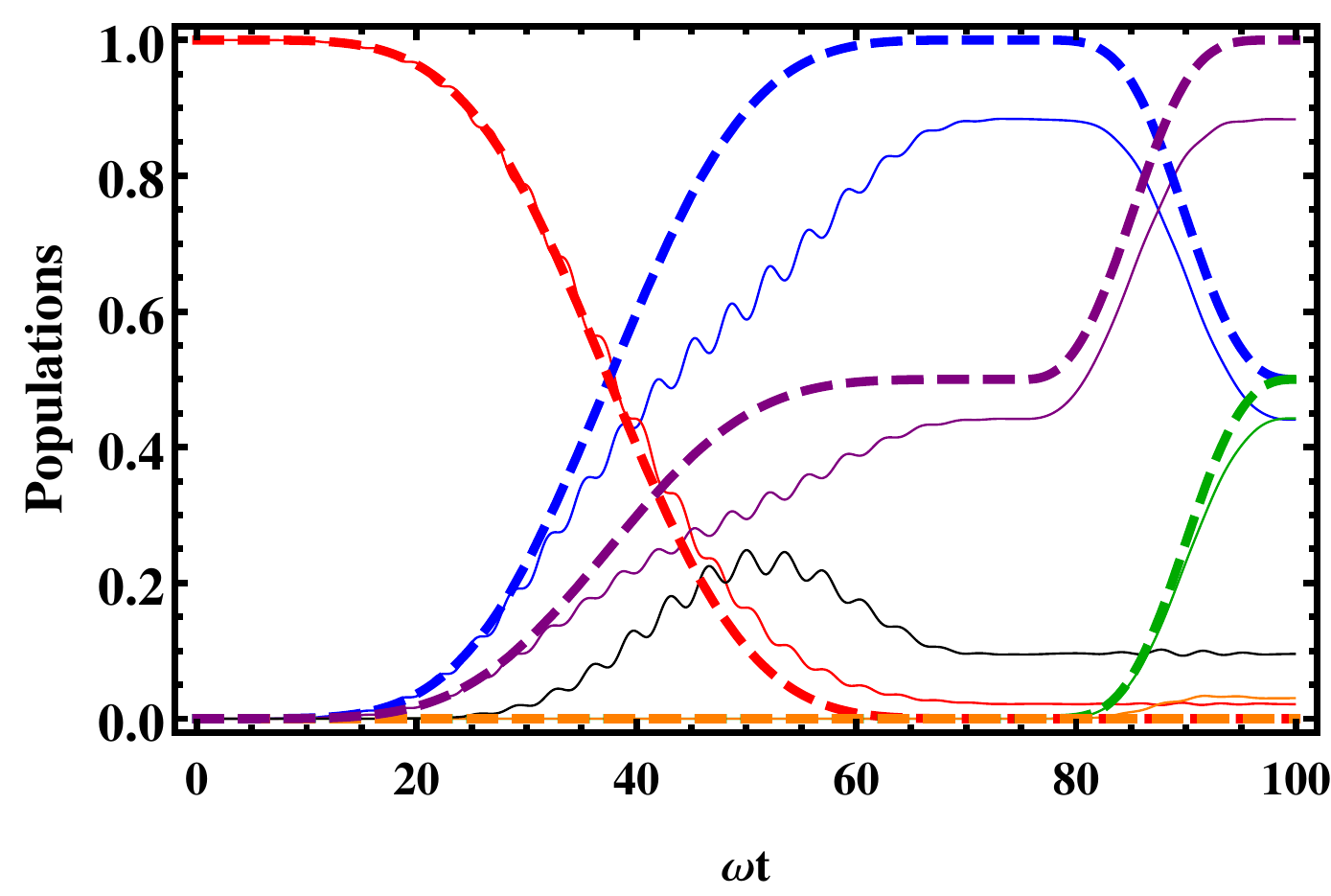}
\hspace{0.5cm}
(d)\includegraphics[angle=0,width=0.43\linewidth]{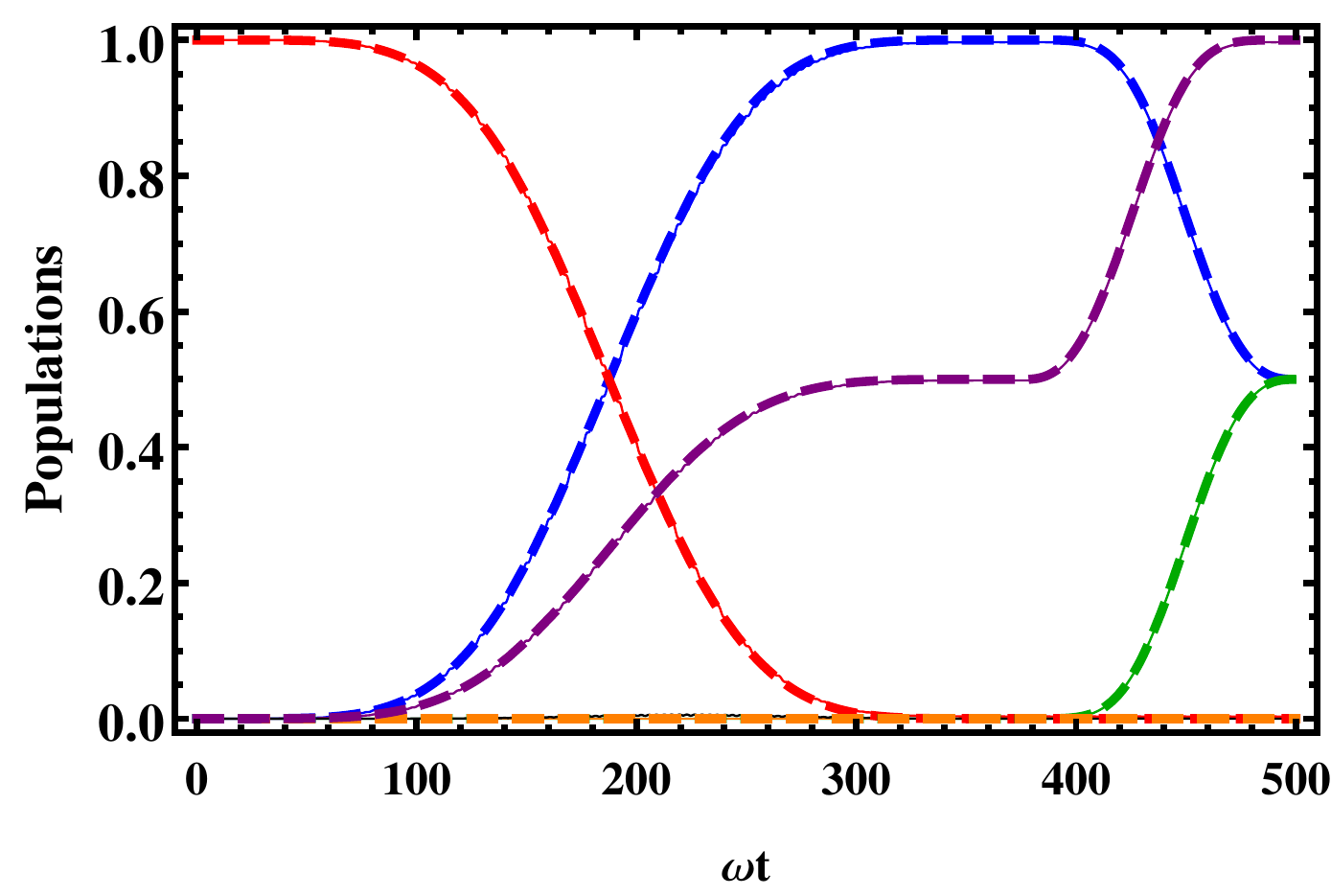}

\end{center}
\caption{Populations against time calculated using the four--level approximation (dashed lines) and the full Schr\"{o}dinger equation (solid lines) with $V_0=3 \hbar \omega$ for
the polynomial process with (a) $T=100 \omega^{-1}$ and
(b) $T=500 \omega^{-1}$ and the piecewise process ($t_S=0.75T$) with (c) $T=100 \omega^{-1}$ and
(d) $T=500 \omega^{-1}$.
Colours correspond to:
  $\left|\braket{\psi(t)}{00}\right|^{2}$ (red),
  $\left|\braket{\psi(t)}{10}\right|^{2}$ (blue),
  $\left|\braket{\psi(t)}{01}\right|^{2}$ (green),
  $\left|\braket{\psi(t)}{11}\right|^{2}$ (orange),
  $\left|\braket{\psi(t)}{-}\right|^{2}$ (purple), and
  populations of higher levels, i.e., $1-\sum_{i,j=0}^{1}\left|\braket{\psi(t)}{ij}\right|^{2}$ (black). \label{populations}}  
\end{figure}

The results of the numerical simulation of both schemes are shown in figure \ref{populations}, together with the ideal populations based on the four--level Hamiltonian in \eqref{H_4L}.
Using the polynomial scheme, even for a short total time $T=100 \omega^{-1}$ (figure \ref{populations}(a)), the final population in the desired state is already greater than $90\%$, with about $5\%$ of population leaking to states outside of the four--level model.
For a longer total time $T$ (figure \ref{populations}(b)), the agreement between the four--level Hamiltonian and the full dynamics is almost perfect, ending up with nearly $100\%$ in the desired state.

Similarly for the piecewise scheme, the dynamics for a short total time (figure \ref{populations}(c)) leads to oscillations and a non--perfect population of the target state, and approximately a $10\%$ population of higher lying states.
However, for longer $T$ (figure \ref{populations}(d)), the final fidelity is nearly $100\%$.
Note that since the second pulse $\Omega_\rho$ in this scheme does not require the rotating wave approximation, it is beneficial to give the first pulse a longer duration.
Hence the choice of $t_S=0.75T$.

\begin{figure}[t]
\begin{center}

(a)\includegraphics[angle=0,width=0.45\linewidth]{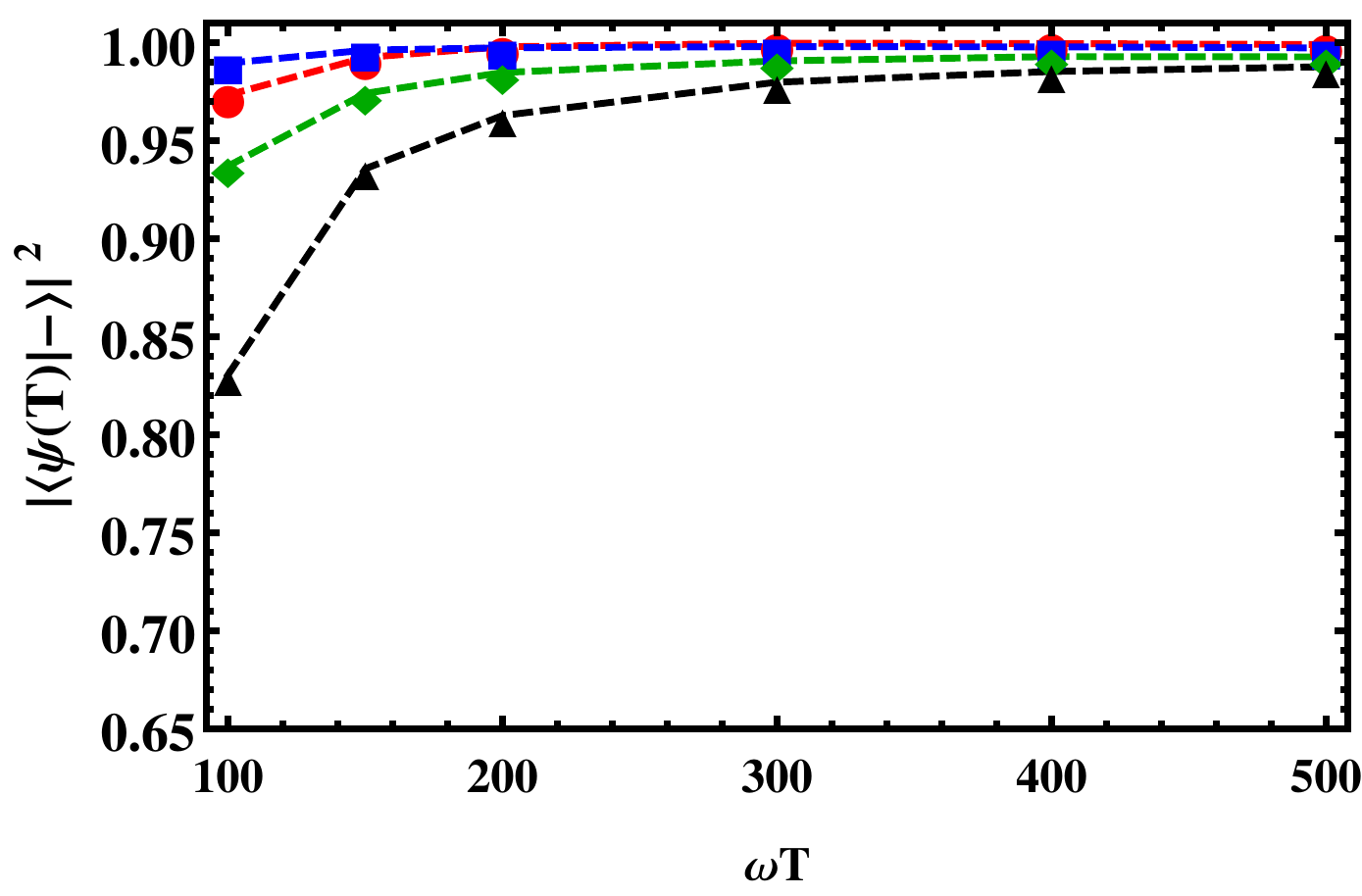}
\hspace{0.1cm}
(b)\includegraphics[angle=0,width=0.45\linewidth]{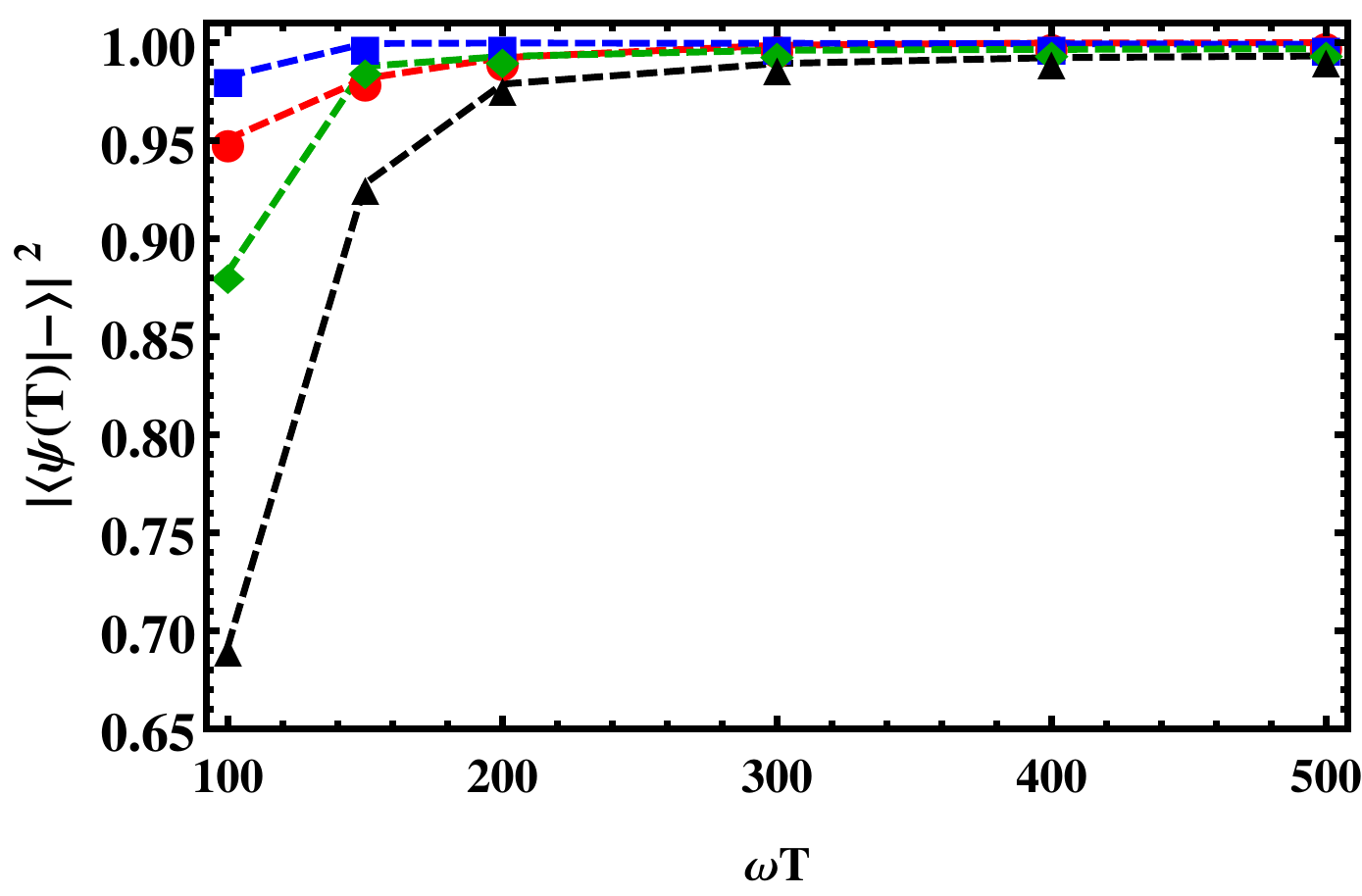}
\end{center}
\caption{Fidelity $\left|\braket{\psi(T)}{-}\right|^2$ against total time T for different lattice depths $V_{0}$ for a fixed trapping frequency $\omega$.
Points joined with lines:
$V_0=2\hbar\omega$ (red circles),
$V_0=2.5\hbar\omega$ (blue squares),
$V_0=3 \hbar\omega$ (green diamonds) and
$V_0=3.5\hbar\omega$ (black triangles);
(a) polynomial scheme,
(b) piecewise scheme ($t_S=0.75T$).
\label{fidelity_plots}
}  
\end{figure}

The fidelity of both schemes for different total times $T$ and different lattice depths $V_0$ is shown in figure \ref{fidelity_plots};
the lattice constant $2\ell = \pi/k$ is varied in such a way that the trapping frequency
$\omega=\sqrt{\frac{2V_0k^{2}}{m}}$ is kept fixed. 
From this we can see again how for a larger $T$ we achieve higher fidelities, which is consistent with the rotating wave approximation and the slowly varying shaking amplitude approximation becoming more valid. We can also see that the fidelities decrease for deeper lattices because as the well becomes deeper it becomes more harmonic and hence has equally spaced energy levels. This leads to resonant coupling to higher energy levels (see \ref{app2} for details).

\begin{figure}[t]
\begin{center}
(a)\includegraphics[angle=0,width=0.45\linewidth]{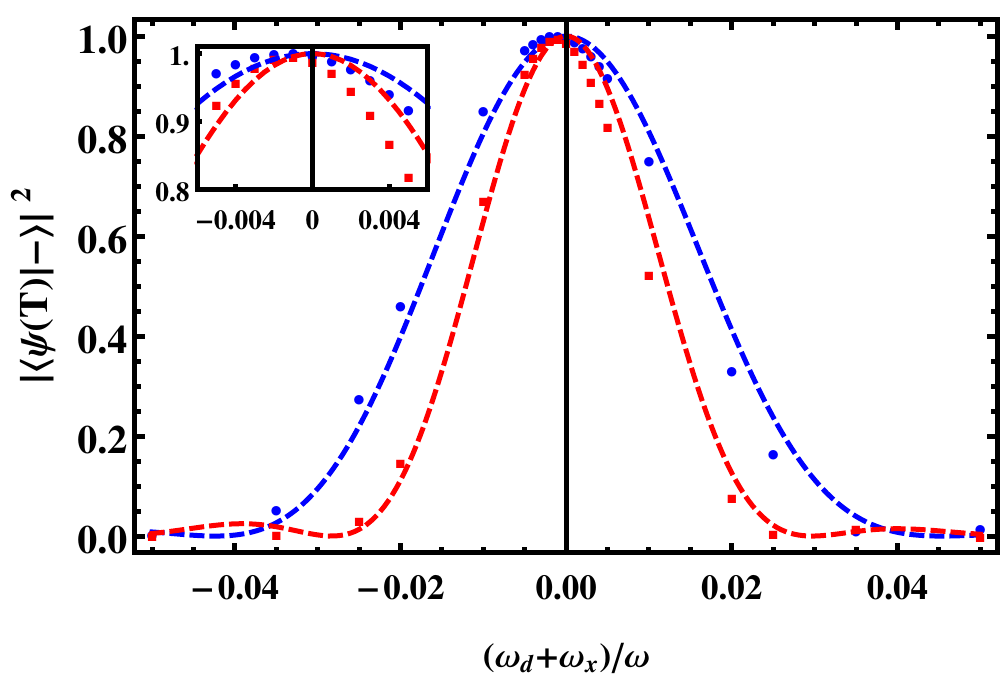}
(b)\includegraphics[angle=0,width=0.45\linewidth]{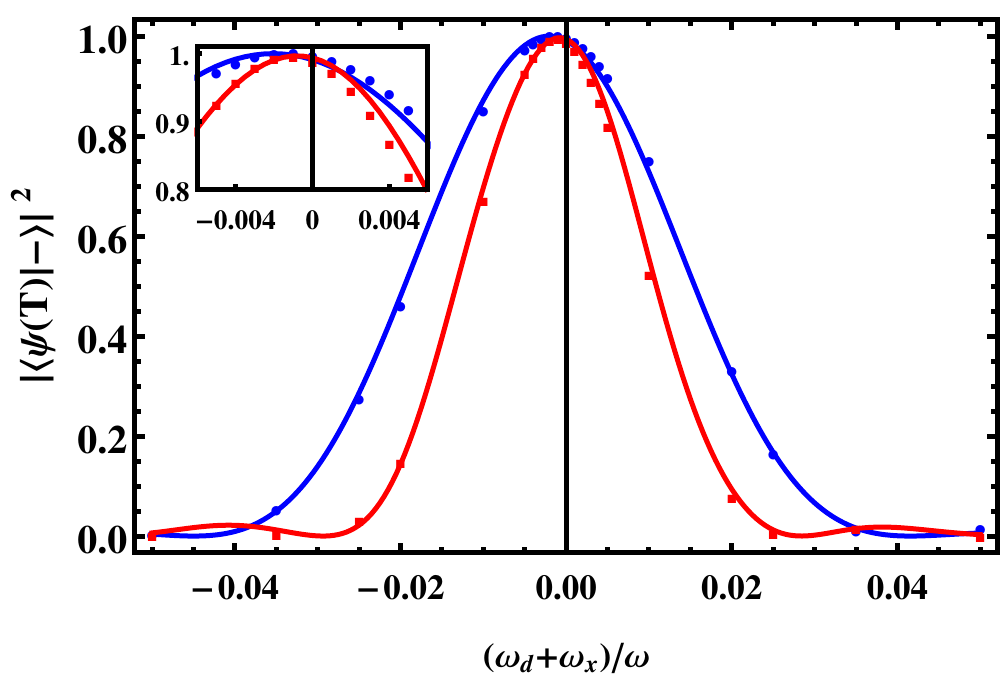}

\end{center}
\caption{Fidelity $\left|\braket{\psi(T)}{-}\right|^2$ against the deviation from resonant shaking $(\omega_{x}+\omega_{d})/\omega$ for $V_0=3\hbar\omega$ and $T=300\omega^{-1}$ (resonant shaking corresponds to $\omega_{x}=-\omega_{d}$).
Polynomial scheme (red) and piecewise scheme with $t_S=0.75T$ (blue).
Points correspond to the full Schr\"{o}dinger equation, dashed lines to the $4$--level model and lines to the $6$--level model \eqref{six_level_H}.
\label{resonance_peak}}  
\end{figure}

In the following, we want to examine the stability of the schemes.
In figure \ref{resonance_peak}(a), we show the resonance curve for both processes, i.e., the fidelity against the detuning of the shaking frequency with respect to the frequency difference of the first two levels. We compare the four--level model(not assuming $\omega_{x}=-\omega_{d}$) against the full Schr\"{o}dinger equation dynamics. As expected, one achieves high fidelity when the shaking frequency is on resonance. Perhaps surprisingly one can note that the highest fidelity of the full dynamics is achieved for a slightly off resonant shaking frequency. This is not true in the four--level model, as the corresponding curves have their maximum at resonance.
The reason for this is the presence of an off resonant coupling to the state $\ket{20}$(which is not present in the four-level model). By slightly increasing the detuning of $\Omega_{x}$ with respect to the $\ket{00} \leftrightarrow \ket{10}$ transition, an even greater detuning in the coupling between $\ket{10}$ and $\ket{20}$ is created, leading to less leakage to these higher states. We can verify this by considering a six--level model (see \ref{app2}), which can be seen to agree with the full Schr\"{o}dinger equation dynamics (see figure \ref{resonance_peak}(b))

\begin{figure}[t]
\begin{center}
\includegraphics[angle=0,width=0.5\linewidth]{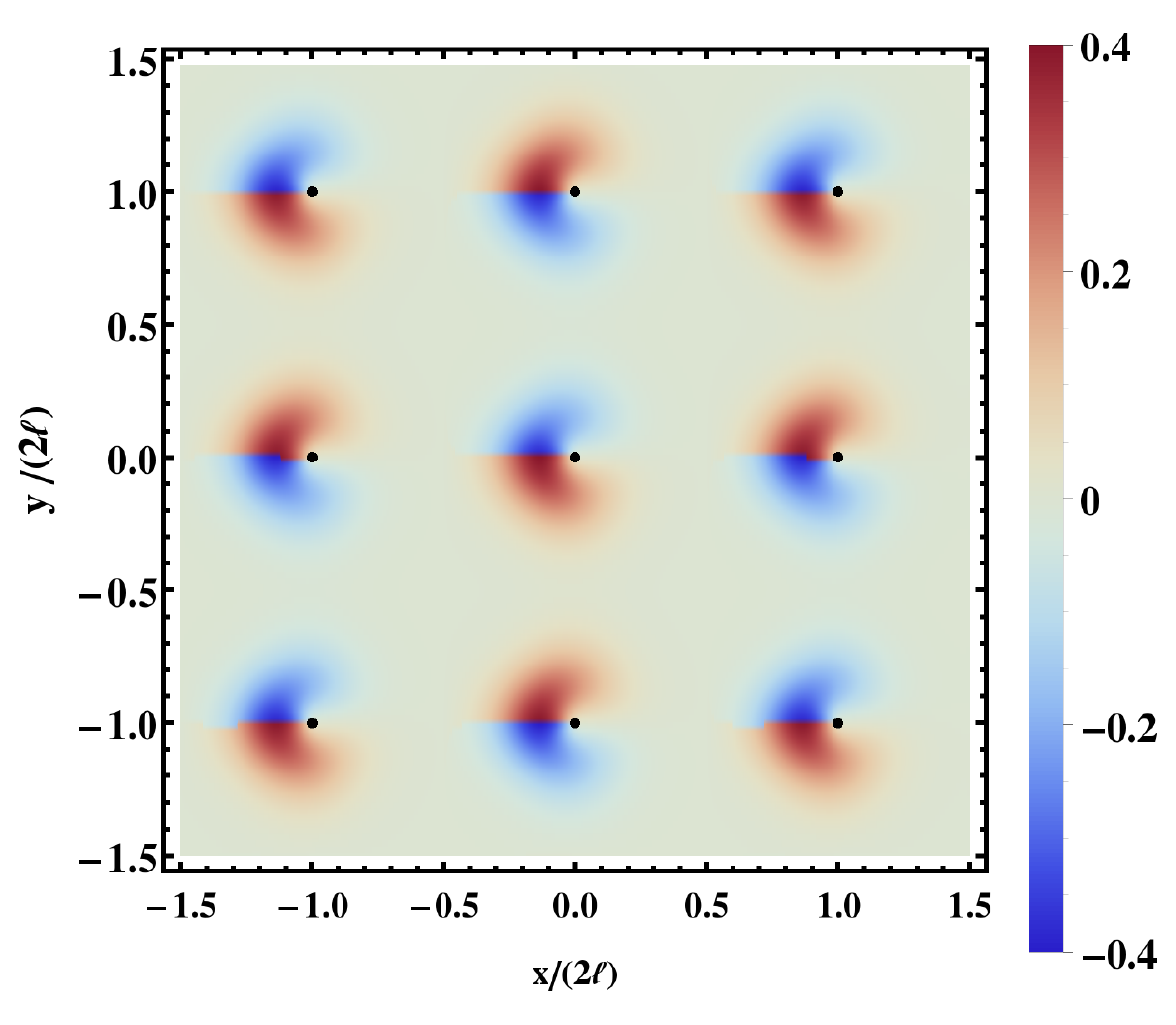}
\end{center}
\caption{Final state after applying the piecewise process with $V_0=3\hbar\omega$, $T=300\omega^{-1}$ and $t_S=0.75T$.
Shown is $\fabs{\Psi (x,y,T)}\cdot\mbox{arg}\left[\Psi(x,y,T)\right]$, with
the black dots indicating the minima of the lattice wells.\label{fig_entanglement}
}  
\end{figure}

Finally, we remark once again that in the case of more lattice sites, each containing a single atom, the schemes would result in the pattern in figure \ref{fig_state_diagram}. As a brief aside, we now consider a single atom whose initial state is now a superposition of all ground states of all $9$ wells of a $3 \times 3$ lattice; the single atom is de--localised across the entire lattice.
Applying here the piecewise shaking scheme, one reaches the final state represented in figure \ref{fig_entanglement}.
It can be clearly seen that a checkerboard pattern of left- and right-handed angular momentum states is produced, similar to figure \ref{fig_state_diagram}.
Note that we have adjusted the (physical irrelevant) global phase such that the branch cut is horizontal in this representation of the wave function.
In this case, we have produced a final state for a single atom in which its position is entangled with the sign of the angular momentum in each well.

\section{Experimental considerations \label{exp_sec}}


There are several options for experimentally implementing such a system depending on how one creates the two counter propagating beams for each direction.
%
One option is to use a beam and a retro-reflecting mirror, in which case one can induce the shaking by mounting the mirror on a piezo-electric actuator which will then oscillate according to $r_{x}(t)$~\cite{ivanov_2008, zenesini_2009, arimondo_2012}.
In the case where the beam is split in two, one can introduce a small frequency difference $\Delta \nu(t)$ between the beams by using acousto-optic modulators to make the lattice move with a velocity $\Delta \nu(t) \lambda/2$, where $\lambda$ is the wavelength of the laser~\cite{sias_2008,arimondo_2012}. The shaking is then given by $r_{x}(t)=\frac{\lambda}{2}\int_{0}^{t}\Delta \nu(\tau)d\tau$.

Parameter values of $V_0/(\hbar \omega)=3$ and $\omega T=300$ could for example be reached using $^{133}$Cs atoms with $\lambda=1064 \, \text{nm}$ lasers and a lattice depth of $36 E_{r}$, where $E_{r}=\frac{\hbar^{2}k^{2}}{2 m}$ is the recoil energy.
The shaking frequency required would be $\omega_{d}/(2\pi) \approx 14\,\text{kHz}$ and the total time required for the operation would be $T \approx 3 \, \text{ms}$.
%
%
Under the assumption that $V_0 \gg E_{r}$ (i.e. that the well is deep), one can approximate the ground state tunnelling rate $J_{0}$ as~\cite{zwerger_2003, bloch_2008}
\begin{eqnarray}
J_{0} \approx \frac{4 E_{r}}{\sqrt{\pi}} \left(\frac{V_0}{E_{r}}\right)^{3/4} e^{-2 \sqrt{V_0/E_{r}}}.
\end{eqnarray}
For our scheme to work, the operation must be performed much faster than this tunnelling time, i.e, we want $T \ll \hbar /J_{0} \approx 589 \, \text{ms}$ for the parameter values above. If one calculates the tunnelling rates using exact band structure calculations \cite{isacsson_2005}, one obtains a ground state tunnelling time of $\hbar /J_{0} \, \approx 600 \text{ms}$ and an excited state tunnelling time of $\hbar /J_{1} \approx \, 17 \text{ms}$. Being in the Mott insulator ground state corresponds to a potential depth of about $22 E_{r}$~\cite{mott_fluid_cs}. Being in the Mott state for both the ground state and the first excited state will not be affected by the shaking, as it has been shown both theoretically~\cite{tunnelling_renorm} and experimentally~\cite{tunnelling_renorm_exp} that the shaking effectively reduces the tunnelling strength to the neighbouring wells. In addition, the anharmonic nature of the potential inhibits first order decay processes ~\cite{isacsson_2005} .

\section{Conclusions \label{discuss}}


We have developed two schemes to prepare an exotic lattice state, namely a staggered order angular momentum state,
starting from a Mott insulator state in an optical lattice.
Both of these are using shaking of the optical lattice together with a modulation of the interference term.
The flexibility of the invariant--based approach presented in this paper makes it possible to extend this research in multiple directions.
For instance, one could further optimise the scheme to combat the most relevant errors in a given experimental implementation \cite{ruschhaupt_2012}. 


It is also possible to extend this idea to prepare a similar state with higher angular momentum per lattice site or by including the $\Omega_{y}$ term to prepare a state with equal angular momentum per lattice site. If the shaking process is applied to a single de--localised atom, the final state can be seen as an entangled state where the well position is entangled with the sign of the angular momentum.

\section*{Acknowledgements}
This work has received financial support from Science Foundation Ireland under the International Strategic Cooperation Award Grant No. SFI/13/ISCA/2845 and the Okinawa Institute of Science and Technology Graduate University.
We are grateful to David Rea for useful discussion and commenting on the manuscript.

\appendix

\section{Transformation into lattice frame\label{app}}

To transform our Hamiltonian in the lab frame,
\begin{eqnarray}
H_\textrm{lab}(t)=\frac{\vec{p}^{\; 2}}{2m}+V(\vec{r}-\vec{R}_{0}(t),t) ,
\end{eqnarray}
to the lattice frame we follow the procedure outlined in~\cite{arimondo_2012}. 
The relationship between the two Hamiltonians is given by a unitary transformation $\cU$,
\begin{eqnarray}
H_\textrm{lattice}(t)=\cU H_\textrm{lab} \cU^{\dagger}-i \hbar \cU\partial_{t}\cU^{\dagger},
\end{eqnarray}
which can be expressed as three separate unitary operators $\cU=U_{3}U_{2}U_{1}$.
These are a translation operator,
\begin{eqnarray}
U_{1}=\exp{\left[\frac{i}{\hbar}\vec{R}_{0}(t)\vec{p}\right]} ,
\end{eqnarray}
a momentum shift operator,
\begin{eqnarray}
U_{2}=\exp{\left[-\frac{i}{\hbar}m\dot{\vec{R}}_{0}(t)\vec{r}\right]} ,
\end{eqnarray}
and an operator that removes a time--dependent energy shift from the Hamiltonian,
\begin{eqnarray}
U_{3}=\exp{\left[-\frac{i}{\hbar}\frac{m}{2}\int_{0}^{t}dt'\dot{\vec{R}}_{0}(t')^{2}\right]}.
\end{eqnarray}
From this we arrive at the Hamiltonian in the lattice frame,
\begin{eqnarray}
H_\textrm{lattice}(t)=\frac{\vec{p}^{\; 2}}{2m}+V(\vec{r},t)+m \ddot{\vec{R}}_{0}(t) \vec{r}
\end{eqnarray}
We impose that $\vec{R}_{0}(0) = \vec{R}_{0}(T) = 0$ and $\dot{\vec{R}}_{0}(0) = \dot{\vec{R}}_{0}(T) = 0$, such that $\cU$ becomes the identity (up to a global phase) at the initial and final times.

\section{Six--level approximation \label{app2}}

If one were to include more levels to approximate the Hamiltonian \eqref{H_lattice_frame}, the natural choice would be $\ket{20}$ and $\ket{02}$.
A six--level Hamiltonian to describe our system can be obtained following a derivation similar to the one presented in \sref{four_level_approx},
but using the unitary operator
\begin{eqnarray}
U(t)&=&e^{-i \omega_{10}t}\ket{10}\bra{10}+e^{-i( \omega_{10}+\omega_{x})t}\ket{00}\bra{00}+e^{-i( \omega_{10}+\omega_{x}-\omega_{y})t}\ket{01}\bra{01}
\nonumber \\ 
&& +e^{-i \omega_{11}t}\ket{11}\bra{11}+ e^{-i \omega_{20}t}\ket{20}\bra{20}+e^{-i \omega_{02}t}\ket{02}\bra{02} 
\end{eqnarray}
and setting $\Omega_{y}=0$. One then arrives at the Hamiltonian
\begin{eqnarray}
H_{6L} = \frac{\hbar}{2}
\left(
\begin{array}{cccccc}
0 & \Omega_{x}\theta_{x}^{-} & \Omega_\rho & 0 & \delta_{1} & 0\\

\Omega_{x} \theta_{x}^{+} & -2\left( \omega_{d}+\omega_{x}\right) & 0 & \Omega_\rho e^{ i \left(\omega_{x}-\omega_{d}\right)t}& 0 & 0\\

\Omega_\rho & 0 & 0 & \Omega_{x} \theta_{x}^{+} & 0 & 0\\

0 & \Omega_\rho e^{-i \left(\omega_{x}-\omega_{d}\right)t} & \Omega_{x} \theta_{x}^{-} & 0 & \delta_{2}  & \delta_{2}\\

\delta_{1}^* & 0 & 0 & \delta_{2}^* & 0 & 0 \\

0 & 0 & 0 & \delta_{2}^* & 0 & 0
\end{array}
\right)
\label{six_level_H}
\end{eqnarray} 
in the ordered basis $\{\ket{10}, \ket{00}, \ket{01}, \ket{11}, \ket{20}, \ket{02}\}$, where
\begin{eqnarray}
\theta_{x}^{\pm} &=& \left(\frac{\omega_{x}}{\omega_{d}}\right)^{2}\left(1+e^{\pm 2 i \omega_{x} t}\right) , \\
\delta_{1} &=& \left[\int_{-\ell}^{\ell} \Gamma_{2}(x)x\Gamma_{1}(x)dx\right]\gamma_{1}^{-1}
\Omega_{x} e^{i\left(\omega_{10}-\omega_{20}+\omega_{x}\right)t}\theta_{x}^{-} , \\
\delta_{2} &=& \left[\int_{-\ell}^{\ell} \Gamma_{2}(x)\sin(k x)\Gamma_{1}(x)dx\right] \frac{1}{\sqrt{\gamma_{2}}}
\Omega_\rho e^{-i (\omega_{20}-\omega_{11}) t}.
\end{eqnarray}

One can see that for deep (i.e. harmonic) potential wells $\omega_{10}-\omega_{20}=-\omega_{d}$ and $\omega_{20}=\omega_{11}$.
For $\omega_{x}=-\omega_{d}$ and in the rotating--wave approximation, one gets
\begin{eqnarray}
H_{6L} = \frac{\hbar}{2}\left(
\begin{array}{cccccc}
0 & \Omega_{x} & \Omega_\rho & 0 & \sqrt{2} \Omega_{x} & 0\\

\Omega_{x} & 0 & 0 & 0& 0 & 0\\

\Omega_\rho & 0 & 0 & \Omega_{x} & 0 & 0\\

0 & 0 & \Omega_{x} & 0 & \sqrt{2}\Omega_\rho  & \sqrt{2} \Omega_\rho \\

\sqrt{2} \Omega_{x}  & 0 & 0 & \sqrt{2}\Omega_\rho & 0 & 0 \\

0 & 0 & 0 & \sqrt{2} \Omega_\rho & 0 & 0
\end{array}
\right) .
\end{eqnarray} 
This clearly shows that for deep lattices a strong resonant coupling to levels $\ket{20}$ and $\ket{02}$ exists, and therefore the four--level approximation becomes invalid in this limit.

\end{document}